\documentclass[pre,twocolumn,a4paper,superscriptaddress]{revtex4-1}
\usepackage[UKenglish]{babel} 
\usepackage{amsmath}
\usepackage[normalem]{ulem}
\usepackage{amssymb}
\usepackage{graphicx}
\usepackage{dcolumn}
\usepackage{bm}
\usepackage{hyperref}
\usepackage{graphicx}
\usepackage{xcolor}
\usepackage{amstext}
\usepackage{amsfonts}
\usepackage{times}
\usepackage{bbold}

\newcommand{\avg}[1]{\left< #1 \right>} % for average
\newcommand{\ket}[1]{\left| #1 \right>} % for Dirac bras
\newcommand{\bra}[1]{\left< #1 \right|} % for Dirac kets
\newcommand{\braket}[2]{\left< #1 \vphantom{#2} \right|
 \left. #2 \vphantom{#1} \right>} % for Dirac brackets
\newcommand{\W}{\mathcal{W}}
\newcommand{\tr}{\text{tr}}
\newcommand{\dt}{\mbox{\bf d}_t}

\AtEndDocument{\clearpage}

\begin{document}

\title{Quantum thermodynamics of general quantum processes}

\author{Felix Binder}
\email{felix.binder@physics.ox.ac.uk}
\affiliation{Clarendon Laboratory, Department of Physics, University of Oxford, Oxford OX1 3PU, United Kingdom}

\author{Sai Vinjanampathy}
%\email{sai@quantumlah.org}
\affiliation{Center for Quantum Technologies, National University of Singapore, 3 Science Drive 2, 117543 Singapore, Singapore}

\author{Kavan Modi}
\affiliation{School of Physics, Monash University, Clayton, Victoria 3800, Australia}

\author{John Goold}
\affiliation{The Abdus Salam International Centre for Theoretical Physics, 34151 Trieste, Italy}

\date{\today}

\begin{abstract}
 Accurately describing work extraction from a quantum system is a central objective for the extension of thermodynamics to individual quantum systems. The concepts of work and heat are surprisingly subtle when generalizations are made to arbitrary quantum states. We formulate an operational thermodynamics suitable for application to an open quantum system undergoing quantum evolution under a general quantum process by which we mean a completely-positive and trace-preserving map. We derive an operational first law of thermodynamics for such processes and show consistency with the second law. We show that heat, from the first law, is positive when the input state of the map majorises the output state. Moreover, the change in entropy is also positive for the same majorisation condition. This makes a strong connection between the two operational laws of thermodynamics.
\end{abstract}

\maketitle

%Thermodynamics as a theory owes its birth primarily to the capitalist doctrines of the 19th century~[K. Marx and F. Engels, Das Kapital: Kritik der politischen O ̈konomie,Vol.3(O.Meissner,1904)]

%*******************************************
%*******************************************
\section{Introduction} The laws of thermodynamics were forged in the furnaces of the industrial revolution, as engineers and scientists refined their picture of energy, studying {\it heat} and its interconversion to mechanical {\it work} with a view to powering the mines and factories of this new era of human endeavor. Followed by the development of statistical mechanics at the change of the centuries \cite{gibbs1902}, far from its pragmatic inception, thermodynamics is now a theory with a remarkable range of applicability, successfully describing the properties of macroscopic systems ranging from refrigerators to black holes \cite{universeandrefrig}. 

Moving on to the 21st century with both industrial and electronic revolutions behind us, technological development is pushing towards and beyond the microscopic scale. With a view to devices operating at a scale where quantum mechanical laws become important we may ask whether the solid grounds of thermodynamics might be challenged, not only by the lack of a thermodynamic limit, but also by the intrinsic uncertainty synonymous with this domain. It comes as no surprise that there has been a concerted effort to understand how the laws of thermodynamics generalize to arbitrary quantum systems~\cite{mahler} both at and away from equilibrium. Such laws will aid in better understanding the relationship between quantum and statistical mechanics, extend our predictability for out-of-equilibrium systems and aid the design of efficient controls for thermal machines.

An important question relating to the extension of the first law of thermodynamics into such a regime is to ask ``to what extent do the concepts of \emph{work} and \emph{heat} extend to quantum systems?" This is an avenue of research that has been open for several decades~\cite{mahler, lindblad, *janet, *Skrzypczyk}. Without severe assumptions regarding the set of allowed quantum states, coupling strengths and bath-properties it has so far remained a difficult question without any satisfactory general answer. Nevertheless it is central for the formulation of a concrete theory of {\it quantum thermodynamics} of both equilibrium and non-equilibrium systems. Some important steps have been made toward providing an answer, such as the formulation of quantum fluctuation relations \cite{mrev, *esposito}, information theoretical approaches \cite{delRio2011, *Dahlsten2011, *Brandao2011, *Horodecki2013, *Aberg2013} (see~\cite{gour} for an overview), and some combination of the two \cite{landauer}. Finally, central to the work presented here is a work extraction formalism for non-passivity of quantum states \cite{Allahverdyan}. Despite the range of approaches a more general picture for the thermodynamics of general quantum evolutions is far from clear. 

In this letter, we take an operational approach to characterising the energy change of an open quantum process described by a \emph{completely-positive trace-preserving} (CPTP) map. Such maps are ubiquitous in modern quantum physics and arguably the most encompassing generic description available for quantum processes (i.e. all processes that can be described by coupling to an initially uncorrelated ancilla, joint unitary evolution, and tracing out over the ancilla). In the context of this paper we will consequently refer to evolutions that are CPTP as 'general quantum processes'. The results presented here rely on processes being both completely-positive and also trace-preserving but are not contingent on a specific description (for instance, in terms of Kraus operators) and the maps may be thought of as an input-output formalism for quantum states. In analogy to the first law of thermodynamics we discuss work done, extractable work, and heat. The concepts of ergotropy and adiabatic work allow us to state our main result: an operational first law for general quantum processes. We show that our operational first law is in agreement with the widely used Hatano-Sasa version of the second law for CPTP maps \cite{Alicki1979, Spohn, *hatano2001steady, *sagawa2013second} by explicitly stating the Clausius inequality for unital and thermal maps. We then show that both operational heat and the change in von Neumann entropy are positive when the input state of the map majorises the output state.

%*******************************************
%*******************************************
\section{Thermodynamics of quantum systems}
The first law of thermodynamics states that the internal energy change in a thermodynamic process can be split into two contributions -- work and heat: $d E=\delta Q+\delta W$. For a general quantum system, the internal energy at time $t$ is $E(t)= \tr[\rho(t) H(t)]$, implying that the change in the internal energy $dE$ depends only on the end points. Heat and work, on the other hand, are \emph{path-dependent} -- hence the different notation for the `differentials'. As illustration we may consider the heat expended when pushing a piston into a cylinder filled with gas: It depends not only on the initial and final positions of the piston but also on how fast it is pushed. Using the time-derivative of the internal energy the following two expressions are motivated~\cite{Alicki1979}:
\begin{align}
\delta W =& \tr\left[\rho(t) \dt H(t) \right] dt \notag
\quad \mbox{and} \\
\delta Q =&\tr\left[H(t) \dt \rho(t) \right]dt \label{eq:workheat},
\end{align}
with $\dt:=d/dt$. Integrated over a specific evolution this yields (average) values for heat and work
\begin{align}
\Delta E =& \int_0^{\tau} \delta W + \int_0^{\tau} \delta Q = \avg{W} + \avg{Q} \notag\\
=&~ \tr [\rho(\tau)H(\tau)]-\tr [\rho(0)H(0)].
\label{eq:1stlaw_integral}
\end{align}
These definitions fit the understanding that heat corresponds to a change in the state and accordingly, entropy, a function of state. For unitary evolution heat vanishes by virtue of the Liouville-von Neumann equation: $\dt \rho(t) = i [H(t), \rho(t)]$. The part corresponding to work, on the other hand, does not relate to a change of the state or its entropy but rather to a change in Hamiltonian.

In general, however, it is not easy to compare work and heat for a general quantum process. This is because the integrands in \autoref{eq:1stlaw_integral} are often neither well-defined nor easy to measure. Only for systems with well-defined descriptors of $\rho(t)$ and $\dt \rho(t)$, do we have a closed form for work and heat. For instance, Markov systems are described in a time local form $\dt \rho(t) = \mathcal{L}[\rho(t)]$, leading to the well-known results~\cite{Alicki1979}. A related situation presents itself when dropping related assumptions about weak coupling, semigroup properties of the quantum channel, or its infinitesimal divisibility. Here we are specifically interested in the regime where such assumptions do not hold. In this sense our approach shares similar obstacles with the popular description of \emph{thermal operations} \cite{Brandao2011, Janzing}. What is different in our scenario is that we do not impose energy-conservation (for instance, in order to derive a second law) but rather ask: Can we meaningfully characterise the energy exchange in a quantum process for which dynamic resolution is not available?

It is clear that trying to recover path-dependent quantities as in \autoref{eq:workheat} would be futile since the `path' (that is, the precise system dynamics) is either unknown or not well defined. In our approach the minimal requirements are the existence of meaningful marginals of the system state and Hamiltonian at both the beginning and the end of the process (Restrictively, one could ask for the system-ancilla state to be separable initially and for the Hamiltonian to be a sum of two local Hamiltonians). However, neither the (marginal) system states nor a system Hamiltonian need to be available and moreover thermodynamically meaningful \emph{during} the time-resolved evolution. Far from being an academic issue, this is a very realistic and practical problem. For instance, in a chemical process the Hamiltonian dynamics as well as the reduced state at all times are generally not known. Such stochastic processes can be described by a CPTP map, which may be thought of as a black box relating an input to an output state.

In this context, work and heat obtain their meaning in an operational sense: Given a general map $M(\rho)$, how much work can be extracted from the output state $\rho'=M(\rho)$ assuming a fully controllable quantum operation? How much energy is wasted (or gained) in the process? The reader may think of the initial state and the map $M$ as free resources in this scenario. In the next two sections we introduce the concepts of \emph{ergotropy} and \emph{adiabatic work} before stating the main result.

%*******************************************
%*******************************************
\section{Ergotropy and cyclic work extraction}
We proceed with a brief review of work extraction in cyclic unitary evolution~\cite{Allahverdyan, lenard, *pusz}: Given a quantum state $\rho$ on a finite-dimensional Hilbert space and a Hamiltonian $H$, we may ask how much work can be extracted via a cyclic unitary process. Cyclicity here means that the system Hamiltonians at the beginning and the end of the process have to be identical, i.e., $H\equiv H(0)=H(\tau)$. For unitary evolution any change in internal energy $\avg{H}$ is due to work. We express the Hamiltonian in its \textit{increasing} spectral decomposition
\begin{equation}
 H:=\sum \epsilon_n \ket{\epsilon_n}\bra{\epsilon_n} \text{, with } \epsilon_{n+1}\geq\epsilon_n \; \forall \; n.
\end{equation}
The state $\rho$, on the other hand, is expressed in its \textit{decreasing} eigen-decomposition
\begin{equation}
 \rho:=\sum r_n \ket{r_n}\bra{r_n}\text{, with } r_{n+1}\leq r_n \; \forall \; n.
\end{equation}
The goal is to transform $\rho$ into a state with lower internal energy, extracting the difference in internal energy in the process. 

After maximal cyclic, unitary work extraction no further work can be extracted and the system ends up in a so called \emph{passive} state $\pi$~\cite{Allahverdyan, lenard, *pusz}. A passive state is unique up to degeneracies in the Hamiltonian~\footnote{Since a passive state's internal energy is fixed even in the presence of degeneracies they have no bearing on the arguments presented here and the reader may think of passive states as being unique with regard to a given Hamiltonian: $\pi$ will here correspond to $H$, $\pi'$ to $H'$.}.
Passive states are diagonal in the Hamiltonian's eigenbasis with decreasing populations for increasing energy levels. That is, a state $\rho$, as defined above, is passive if $\ket{r_n} = \ket{\epsilon_n} \; \forall \; n$. Gibbs states are consequently passive.

The maximum work that can be extracted from a non-passive state $\rho$ with respect to a Hamiltonian $H$ via a cyclic unitary process ($\rho \to \pi$) is called \emph{ergotropy}~\cite{Allahverdyan}:
\begin{align}
\mathcal{W} :=\tr[\rho H -\pi H] = \sum_{m,n}r_m \epsilon_n \left[ \left\vert\braket{\epsilon_n}{r_m}\right\vert^2-\delta_{mn}\right].
\end{align}
Ergotropy is always positive and includes contributions to work extraction from both excitations and coherences. Its relation to quantum correlations has recently been explored in \cite{Hovhannisyan,*fannes,*Campbell}. We may write $\mathcal{W}(\rho,H)$ in order to explicitly state the dependence on the pair $(\rho,H)$ of state and Hamiltonian.

%*******************************************
%*******************************************
\section{Adiabatic work}
Consider now a non-cyclic, unitary process with different initial and final Hamiltonian $H$ and $H':=\sum\epsilon'_n \ket{\epsilon'_n}\bra{\epsilon'_n}$, again with $\epsilon_{n+1}'\geq \epsilon_n'$. If we restrict the change in the Hamiltonian from $H$ to $H'$ to be adiabatic in the quantum sense, i.e., the eigenstates of the Hamiltonian remain eigenstates at each instant, the final state $\pi'$ will be a passive state with respect to $H'$ if the initial state $\pi_m$ was passive with respect to $H$. Since this transformation is unitary there is no heat, and we call the energy change \textit{adiabatic work}:
\begin{gather} \label{eq:Wad}
\avg{W}_{ad}=\tr[\pi' H']-\tr[\pi_m H].
\end{gather}
In the following we will associate this definition with any process, adiabatic or not, that starts with a passive state and preserves its spectrum.

In the case of a general unitary process $(\rho,H)\rightarrow(\rho',H')$ we can combine the ideas of ergotropy and adiabatic work. Since all energy change is work it can be extracted reversibly from the final state if full quantum control is available. The extractable work in $\rho'$ is thus given by 
\begin{equation}
W=\avg{W}_{ad}+\W(\rho',H')-\W(\rho,H). 
\end{equation}
If the initial state is passive we can think of the ergotropy as \emph{deposited work}, or \emph{inner friction} as in \cite{Plastina}, where initial Gibbs states were considered.

%*******************************************
%*******************************************
\section{Energetics of open quantum evolution} We now consider the change in internal energy $E$ due to general quantum evolution, i.e. $\rho \to \rho' =: M(\rho)$:
\begin{gather}
\Delta E=\tr[\rho'  H']-\tr[\rho H],
\end{gather}
where we have labeled the initial and final system Hamiltonian $H$ and $H'$. We may now use the notions of ergotropy and adiabatic work as introduced above to arrive at an operationally meaningful first law of thermodynamics. Defining $\pi_m:=\sum_nr'_n\ket{\epsilon_n}\bra{\epsilon_n}$ we add and subtract $\tr[\pi' H']$, $\tr[\pi_m H']$, and $\tr[\pi H]$ to $\Delta E$ to get
\begin{align}
\Delta E =& \tr[\pi H] -\tr[\rho H] + \tr[\rho'  H'] - \tr[\pi' H'] \notag \\
& + \tr[\pi' H'] - \tr[\pi_m H] + \tr[\pi_m H] - \tr[\pi H].
\end{align}
The first two pairs of terms are simply the ergotropies $-\mathcal{W}(\rho,H) + \mathcal{W}(\rho^\prime,H^\prime) =: \Delta\W$, while the next pair is adiabatic work. Defining operational heat
\begin{equation}
 \avg{Q}_{op}= \tr[\pi_m H] - \tr[\pi H]
\end{equation}
we state the main result:
\begin{gather}
\Delta E=\Delta \mathcal{W} +\avg{W}_{ad} + \avg{Q}_{op} .
\label{eq:main}
\end{gather}
This last equation tells us that the internal energy change in a general quantum process can be split up into a work-like, a heat-like and a third, genuine out-of-equilibrium contribution that equals the difference in ergotropy between initial and final state. In this sense it can be understood as an operational first law of quantum thermodynamics. The definition of heat is justified as the eigenvalues of equilibrium state $\pi_m$ are changed to attain another equilibrium state $\pi$, in analogy to the second term of \autoref{eq:workheat}. This expression of the first law becomes particularly meaningful for processes where the internal energy remains constant but the ergotropy of the state changes -- A conventional description of the first law is inadequate for recognising this change. We illustrate this in the next section.

%*******************************************
\begin{figure}[t]
\centering
\includegraphics[width=0.38\textwidth]{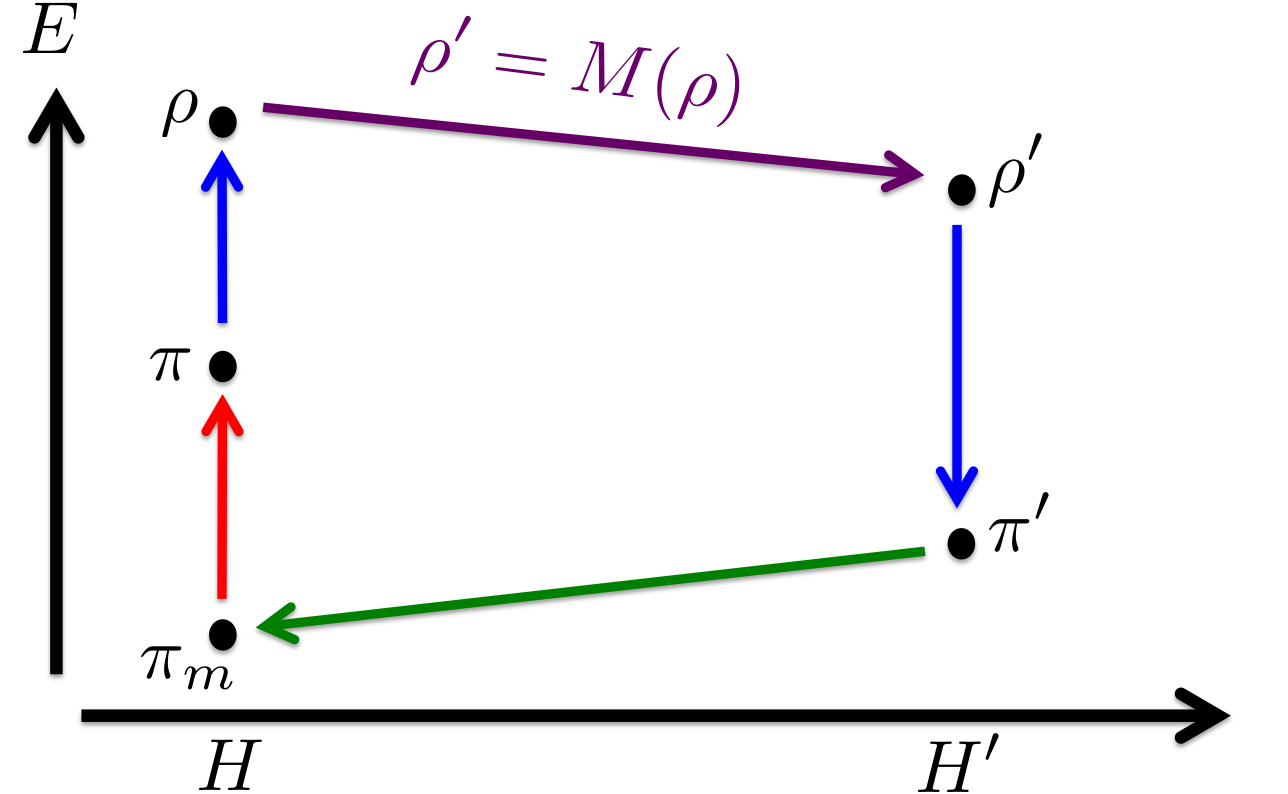}
\caption{(Color online.) The process starts with a non-passive state $\rho$ and ends up in state $\rho'=M(\rho)$. We express this out-of-equilibrium change in internal energy between $\rho$ and $\rho'$ (purple) using a plausible, but not necessarily implemented, reverse process of extracting ergotropy $\rho' \to \pi'$ (blue), equilibrium extraction of adiabatic work $\pi' \to \pi_m$ (green), equilibrium heating $\pi_m \to \pi$ (orange), and finally re-depositing ergotropy $\pi \to \rho$ (blue) to close the loop. The relationship between these quantities is given in \autoref{eq:main}. It is worth noting that whilst the processes in the graph need not be implemented in practice, they are operationally meaningful as illustrated below.
\label{fig:map-nonpassive}}
\end{figure}
%*******************************************

Now, we can interpret the CPTP map as a sequence of fictitious thermodynamic processes as illustrated in \autoref{fig:map-nonpassive}. Going backwards one can extract an amount of work equal to $\mathcal{W}(\rho',H')+\avg{W}_{ad}$. This is the maximal amount that can be extracted unitarily from $\rho'$ ending up back in $H$. After work extraction the state ends up in $\pi_m$. In order to complete the cycle by also resetting the state a heat step is necessary to return the original spectrum of $\rho$. The heat in this process, going from $\pi_m$ to $\pi$, is given by $-\avg{Q}_{op}$. Finally, we restore coherences and excitations that might have been present in $\rho$ by re-inserting $\mathcal{W}(\rho,H)$. This last step is of course only necessary if the initial state was active. For illustration we now provide an example where the initial state is passive.

\subsection*{Example: Initial thermal state}
In a particularily relevant setting we may consider the process to start in a Gibbs state $\tau_\beta=\rho=\pi$. This is a natural setting to consider if, similar to thermal operations, a heat bath at inverse temparature $\beta$ is available as a resource. The setting is the same as before, in \autoref{fig:map-nonpassive}, with $\pi=\rho$. As a consequence, no additional deposition of ergotropy is possible at the end of the cycle. As before the ergotropy gained in going from $\rho'$ to $\pi'$ is the maximum work that can be extracted in a cyclic process with initial state $\rho'$ and reference Hamiltonian $H'$. The maximum work that can unitarily be extracted in a non-cyclic process with final Hamiltonian $H$ is given by $\avg{W}_{max}:=\W'+\avg{W}_{ad}$. Whilst $\W'$, being a genuine out-of-equilibrium quantity, is always positive, $\avg{W}_{ad}$ could also be negative. Requiring the process to finish with $H$ (rather than $H'$) could thus in fact be disadvantaguous for work extraction since, depending on $H$, $\avg{W}_{max}<\W'$ is possible. In the last step (the green arrow in the graph) the transfer of operational heat going from $\pi_m$ to $\pi(=\tau_\beta)$ corresponds to the very practical scenario of thermalisation at the initial temperature $\beta^{-1}$ and Hamiltonian $H$ after maximal work extraction. This concludes the example.
\vspace{\baselineskip}

In summary we note that it has long been convenient to express the properties of out-of-equilibrium systems by using quantities that relate to equilibrium states and hence can be computed in a straight-forward fashion. Furthermore, the equilibrium quantities relate to measurements that can predict the non-equilibrium properties of systems of interest. Similar to other works in statistical mechanics relating to fluctuation theorems \cite{mrev,esposito} we have broken up the out-of-equilibrium energy change into equilibrium quantities adiabatic work and operational heat, and an operational quantity ergotropy. We now relate these results to the second law of thermodynamics.

%*******************************************
%*******************************************
\section{Operational heat and dephasing} We illustrate the meaning of operational heat for non-equilibrium processes. Similar to thermal operations we consider a time-independent system Hamiltonian $H$ and a unitary $V$ on a composite system-ancilla Hilbert space with $[V,H\otimes\mathbb{1}_A]=0$. In this process the total change of the system's internal energy is zero as is the work (due to the Hamiltonian's time-independence). Consequently, the (conventional) heat must also vanish. The ergotropy of the state, however, can change during such a process.

For illustration we first consider an example where system and ancilla are both given by a qubit and the evolution happens according to an interaction Hamiltonian $H_{int}=\sigma_z\otimes\sigma_z$ for $0\leq t\leq\tau$. Starting with a generic state $\rho_0=\left( 
\begin{array}{cc}
p & c\\
c^* & 1-p
\end{array}
\right)$ and an ancilla state $\sigma_0=\left( 
\begin{array}{cc}
\frac{1}{2} & 0\\
0 & \frac{1}{2}
\end{array}
\right)$ the dynamics can be solved exactly:
\begin{equation}
\rho(t) = \left( 
\begin{array}{cc}
p & c \cos 2t\\
c^*\cos 2t & 1-p 
\end{array}
\right)
\end{equation}
It is apparent that whilst the state's internal energy does not change (for a local Hamiltonian in $\sigma_z$-eigenbasis) the ergotropy decreases due to a loss of coherence (periodic revival occurs at times $t=\frac{n}{2}\pi$). According to the main result of this article, the negative change in ergotropy is compensated by an operational heat flow into the system, i.e. $\Delta \mathcal{W}=-\avg{Q}_{op}$.

It can be shown that this holds true more generally for systems of arbitrary dimension and all interaction Hamiltonians -- time independent or not -- that commute with the system Hamiltonian, $[H_{int},H\otimes \mathbb{1}_A]=0$. The populations of the state remain the same but the coherences decrease over time due the open systems dynamics leading to a change in ergotropy which is compensated by operational heat.

Note that dephasing channels such as qubit dynamics governed by a master equation in Linblad form: 
\begin{equation}
 \dot\rho=\gamma(t)(\sigma_z\rho\sigma_z-2\rho)
\end{equation}
are a prominent subset of such dynamics.

With this intuition for the operational heat in mind we now relate the operational first law to an operational second law of thermodynamics.

%*******************************************
%*******************************************
\section{Connecting first and second laws}
Interestingly, a second law for CPTP maps is well-known \cite{Spohn, *hatano2001steady, *sagawa2013second}. In the context of equilibrium thermodynamics, the Clausius inequality states that the thermodynamic entropy of any system and its environment is non-decreasing. For systems in equilibrium, owing to the notions of temperature $\beta^{-1}$, thermodynamic entropy $\Delta S$ and heat $\avg{Q}$ being well defined, the second law can be stated as $\Delta S \geq \beta \avg{Q}$.

To generalize this to the quantum regime, von Neumann entropy, $S(\rho):=\tr[\rho\log(\rho)]$, is considered in the place of thermodynamic entropy (being equivalent for thermal states). The second law for arbitrary states undergoing CPTP evolution is a direct consequence of the fact that relative entropy, defined as \cite{vedral2002role} 
\begin{equation}
S[\rho \Vert \sigma]:=\tr[\rho\log(\rho)-\rho\log(\sigma)],
\end{equation}
obeys contractivity under CPTP maps~\cite{lindblad1975completely}:
\begin{equation}
S[\rho\Vert\sigma] \geq S[M(\rho) \Vert M(\sigma)].
\end{equation}
Since we are interested in the change in entropy $\Delta S:=S(M(\rho)) - S(\rho)$, we have the choice  of a reference state $\sigma$. The obvious choice of $\sigma$ is the fixed point $\mathfrak{e}$ of the map $M$, i.e. $M(\mathfrak{e})=\mathfrak{e}$. Rearranging the contractivity inequality, we arrive at the quantum version of the Hatano-Sasa inequality \cite{Spohn, *hatano2001steady, *sagawa2013second, secondLaw}:
\begin{equation}
 \Delta S\geq -\tr[\{M(\rho)-\rho\}\log(\mathfrak{e})].
\end{equation}

While the first law relates to the partitioning of energy into heat and work, the (Clausius form of the) second law relates only to the increase in entropy. Specifically, the quantum Hatano-Sasa inequality is valid for CPTP evolution where neither heat nor temperature are well defined quantities. Hence, in general it is difficult to verify the internal consistency between a quantum generalization of the first law and a similar generalization of the second law that is applicable to arbitrary CPTP dynamics. However, we establish a relation between the two laws by considering thermal maps.

%*******************************************
\subsection*{Thermal maps}
We call a map thermal if it has a thermal state for a fixed point: $\mathfrak{e}=\tau_\beta=\exp(-\beta\{H-F\})$ at some temperature $\beta^{-1}$, where $F$ is the (Helmholtz) free energy. Such maps, sometimes also called 'Gibbs-preserving maps', are a superset of thermal operations (as is easy to show and further elaborated in \cite{faist}). Consequently, all results related to thermal maps presented here equally apply to the popular set of thermal operations. We remind the reader that all thermal states are passive. In order to make the connection to the second law a cyclic process is considered, i.e., $H=H'$. The input and output states $\rho$ and $\rho'$ are not restricted and can both be out-of-equilibrium. The quantum Hatano-Sasa inequality now reduces to the familiar version of the second law:
\begin{align}
\Delta S=S(\rho')-S(\rho)\geq&-\tr [(\rho'-\rho)\log (\mathfrak{e})] \notag\\
	  =&\beta \,\tr [(\rho'-\rho)(H-F)] \notag\\
	  =&\beta \, (\Delta \mathcal{W}+\avg{Q}_{op}),
\end{align}
with the change in ergotropy playing the role of heat along with $\avg{Q}_{op}$. This restatement of the quantum Hatano-Sasa inequality is interesting in that it lower bounds the entropic change by the sum of two terms, the change in ergotropy and the operational heat which are both measurable and operationally well defined.

%*******************************************
%*******************************************
\section{Majorisation, entropy, and heat}
In order to give a condition for when heat $\avg{Q}_{op}$ is positive (negative) we now introduce the concept of majorisation: A state $\rho$ is said to majorise $\rho'$ (written $\rho \succ \rho'$) if the eigenvalues of the two states satisfy
\begin{gather}
\sum_{m=1}^nr_m \geq \sum_{m=1}^nr'_m \; \forall \; n
\end{gather}
where $\rho' =\sum_m r'_m \ket{r'_m} \bra{r'_m}$ with $r'_{m+1} \ge r'_m$. Note that not all pairs of states obey a majorisation relation -- Some states are incomparable. In those cases we cannot make a statement about operational heat based on the states alone -- It will also depend on the level spacings.

Majorisation provides a sufficient criterion for operational heat to be positive (negative): If $\rho \succ \rho'=M(\rho)$ ($\rho \prec \rho'=M(\rho)$) then $\avg{Q}_{op} \geq 0$ ($\avg{Q}_{op} \leq 0$). Since the eigenvalues of the states do not change during ergotropy extraction $\rho \succ \rho'$ implies $\pi \succ \pi'$. Examining the expression for $\avg{Q}_{op}$ we have:
\begin{align}
\avg{Q}_{op}=&\tr[\pi' H]-\tr[\pi H]
=\sum (r'_n-r_n)\epsilon_n \notag\\
=&\sum_n(\epsilon_{n+1}-\epsilon_n) \sum_{m=1}^n(r_m-r'_m),
\label{eq:difference}
\end{align}
where each term in the last line is positive \footnote{In the same way, adiabatic work is positive when $H' \succ H$. Unlike for operational heat, there is, however, no connection between the majorisation relations for $H$ and $H'$ and the type of map that governs the process.}. 

Moreover, if $\pi \succ \pi'$ then $f(\pi') \ge f(\pi)$ for any Schur-concave function $f$ \cite{Zyczkowski}. The means $S(\rho') \ge S(\rho)$ and therefore $\Delta S \ge 0$. The implication is rather profound: Majorisation guarantees that both the entropy change and operational heat are positive simultaneously. This can be thought of as a version of the second law. 

In the context of equilibrium thermodynamics, the second law guarantees that $\avg{Q} \geq 0$, $\Delta S\geq0$  and that the latter is at least as big as $\beta$ times the former. But in the context of quantum thermodynamics, no such guarantee exists in general. Consequently, one can have cooling transformations that reduce the entropy of the states \cite{secondLaw}. Majorisation strongly restricts the set of allowed transformations to those with positive operational heat and increasing entropy.

%*******************************************
\subsection*{Unital maps}
Unital maps take the maximally mixed state onto itself: $M_u(\mathbf{I}/d)=\mathbf{I}/d$. This simple condition has strong consequences: The quantum Hardy-Littlewood-Polya theorem~\cite{Bengtson} demonstrates that $\rho \succ M_u(\rho)$ for any $\rho$ if $M_u$ is unital. As a consequence of the majorisation arguments above both $\avg{Q}_{op}$ and $\Delta S$ are thus positive for \emph{any} input state $\rho$ and unital map $M_u$.

For all non-unital maps, such as the thermal ones described above, there exists at least one state $\rho$ (the maximally mixed one being a trivial example) that is majorised by the outcome state $\rho'$. In these cases the amount of accessible work increases, i.e., the last term in \autoref{eq:difference} is non-positive and therefore operational heat is less than or equal to zero. The change in entropy for such a process will also be non-positive. The directionality that comes with the second law of thermodynamics is here reflected in the asymmetry between unital and non-unital maps: A heat-like increase of energy (and entropy) of a state only requires a unital map, whilst extraction corresponding to a negative heat-like contribution necessitates a non-unital channel. The representation of such a channel in terms of an ancilla makes clear that the second law is not `violated' here.

In addition to unital maps we may define \emph{anti-unital} maps as those for which any output state majorises the corresponding input state. In such a process the operational heat is always negative.

%*******************************************
%*******************************************

\section{Conclusion}
In summary, we have established a relation that gives an explicit energy balance for all quantum processes that are completely-positive and trace-preserving. Crucially, we have formulated an operational framework for the thermodynamics of open quantum systems. The important feature of this framework is that it relies only on general quantum processes that connect input and output states. Complete-positivity and trace-preservation guarantee that output states are in fact 'physical'. For such processes we have then operationally defined heat and connected it to an operational second law. Both heat and change in entropy are shown to be positive when the input majorises the output, making a strong connection between the operational laws. Furthermore, we have pointed out that the conventional thermodynamic description of quantum processes in terms of projective energy measurements alone does not suffice to capture the change in work value when coherences and excitations of the state are possible. In such cases a change in extractable work (ergotropy) is compensated by an (operational) heat flow into the system, thus giving a concrete meaning to operational heat.

%*******************************************
\begin{acknowledgments}
{\bf Acknowledgments.}	
The authors thank O. Dahlsten and V. Vedral for insightful comments. FB acknowledges funding by the Rhodes Trust. Centre for Quantum Technologies is a Research Centre of Excellence funded by the Ministry of Education and the National Research Foundation of Singapore. KM acknowledges the Templeton Foundation for support. This work was partially supported by the COST Action MP1209. JG and FB thank Kate Clow, the author of the guide book {\it The Lycian way}, a path on which some of the central ideas of this work were conceived.  
\end{acknowledgments}
%*******************************************

\bibliography{paper.bib}

\begin{thebibliography}{35}%
\makeatletter
\providecommand \@ifxundefined [1]{%
 \@ifx{#1\undefined}
}%
\providecommand \@ifnum [1]{%
 \ifnum #1\expandafter \@firstoftwo
 \else \expandafter \@secondoftwo
 \fi
}%
\providecommand \@ifx [1]{%
 \ifx #1\expandafter \@firstoftwo
 \else \expandafter \@secondoftwo
 \fi
}%
\providecommand \natexlab [1]{#1}%
\providecommand \enquote  [1]{``#1''}%
\providecommand \bibnamefont  [1]{#1}%
\providecommand \bibfnamefont [1]{#1}%
\providecommand \citenamefont [1]{#1}%
\providecommand \href@noop [0]{\@secondoftwo}%
\providecommand \href [0]{\begingroup \@sanitize@url \@href}%
\providecommand \@href[1]{\@@startlink{#1}\@@href}%
\providecommand \@@href[1]{\endgroup#1\@@endlink}%
\providecommand \@sanitize@url [0]{\catcode `\\12\catcode `\$12\catcode
  `\&12\catcode `\#12\catcode `\^12\catcode `\_12\catcode `\%12\relax}%
\providecommand \@@startlink[1]{}%
\providecommand \@@endlink[0]{}%
\providecommand \url  [0]{\begingroup\@sanitize@url \@url }%
\providecommand \@url [1]{\endgroup\@href {#1}{\urlprefix }}%
\providecommand \urlprefix  [0]{URL }%
\providecommand \Eprint [0]{\href }%
\providecommand \doibase [0]{http://dx.doi.org/}%
\providecommand \selectlanguage [0]{\@gobble}%
\providecommand \bibinfo  [0]{\@secondoftwo}%
\providecommand \bibfield  [0]{\@secondoftwo}%
\providecommand \translation [1]{[#1]}%
\providecommand \BibitemOpen [0]{}%
\providecommand \bibitemStop [0]{}%
\providecommand \bibitemNoStop [0]{.\EOS\space}%
\providecommand \EOS [0]{\spacefactor3000\relax}%
\providecommand \BibitemShut  [1]{\csname bibitem#1\endcsname}%
\let\auto@bib@innerbib\@empty
%</preamble>
\bibitem [{\citenamefont {Gibbs}(1902)}]{gibbs1902}%
  \BibitemOpen
  \bibfield  {author} {\bibinfo {author} {\bibfnamefont {J.~W.}\ \bibnamefont
  {Gibbs}},\ }\href@noop {} {\emph {\bibinfo {title} {Elementary Principles in
  Statistical Mechanics}}}\ (\bibinfo  {publisher} {Scribner, New York},\
  \bibinfo {year} {1902})\BibitemShut {NoStop}%
\bibitem [{\citenamefont {Goldstein}\ and\ \citenamefont
  {Goldstein}(1993)}]{universeandrefrig}%
  \BibitemOpen
  \bibfield  {author} {\bibinfo {author} {\bibfnamefont {M.}~\bibnamefont
  {Goldstein}}\ and\ \bibinfo {author} {\bibfnamefont {F.~I.}\ \bibnamefont
  {Goldstein}},\ }\href@noop {} {\emph {\bibinfo {title} {The Refrigerator and
  the Universe}}}\ (\bibinfo  {publisher} {Harvard University Press,
  Cambridge},\ \bibinfo {year} {1993})\BibitemShut {NoStop}%
\bibitem [{\citenamefont {Gemmer}\ \emph {et~al.}(2008)\citenamefont {Gemmer},
  \citenamefont {Michel},\ and\ \citenamefont {Mahler}}]{mahler}%
  \BibitemOpen
  \bibfield  {author} {\bibinfo {author} {\bibfnamefont {J.}~\bibnamefont
  {Gemmer}}, \bibinfo {author} {\bibfnamefont {M.}~\bibnamefont {Michel}}, \
  and\ \bibinfo {author} {\bibfnamefont {G.}~\bibnamefont {Mahler}},\
  }\href@noop {} {\emph {\bibinfo {title} {Quantum Thermodynamics}}}\ (\bibinfo
   {publisher} {Springer},\ \bibinfo {year} {2008})\BibitemShut {NoStop}%
\bibitem [{\citenamefont {Lindblad}(1983)}]{lindblad}%
  \BibitemOpen
  \bibfield  {author} {\bibinfo {author} {\bibfnamefont {G.}~\bibnamefont
  {Lindblad}},\ }\href@noop {} {\emph {\bibinfo {title} {Non-Equilibrium
  Entropy and Irreversibility}}}\ (\bibinfo  {publisher} {Reidel, Lancaster},\
  \bibinfo {year} {1983})\BibitemShut {NoStop}%
\bibitem [{\citenamefont {Anders}\ and\ \citenamefont
  {Giovannetti}(2013)}]{janet}%
  \BibitemOpen
  \bibfield  {author} {\bibinfo {author} {\bibfnamefont {J.}~\bibnamefont
  {Anders}}\ and\ \bibinfo {author} {\bibfnamefont {V.}~\bibnamefont
  {Giovannetti}},\ }\href@noop {} {\bibfield  {journal} {\bibinfo  {journal}
  {New J. Phys.}\ }\textbf {\bibinfo {volume} {15}},\ \bibinfo {pages} {033022}
  (\bibinfo {year} {2013})}\BibitemShut {NoStop}%
\bibitem [{\citenamefont {Skrzypczyk}\ \emph {et~al.}(2014)\citenamefont
  {Skrzypczyk}, \citenamefont {Short},\ and\ \citenamefont
  {Popescu}}]{Skrzypczyk}%
  \BibitemOpen
  \bibfield  {author} {\bibinfo {author} {\bibfnamefont {P.}~\bibnamefont
  {Skrzypczyk}}, \bibinfo {author} {\bibfnamefont {A.}~\bibnamefont {Short}}, \
  and\ \bibinfo {author} {\bibfnamefont {S.}~\bibnamefont {Popescu}},\
  }\href@noop {} {\bibfield  {journal} {\bibinfo  {journal} {Nat. Commun.}\
  }\textbf {\bibinfo {volume} {5}},\ \bibinfo {pages} {4185} (\bibinfo {year}
  {2014})}\BibitemShut {NoStop}%
\bibitem [{\citenamefont {Campisi}\ \emph {et~al.}(2011)\citenamefont
  {Campisi}, \citenamefont {H\"{a}nggi},\ and\ \citenamefont {Talkner}}]{mrev}%
  \BibitemOpen
  \bibfield  {author} {\bibinfo {author} {\bibfnamefont {M.}~\bibnamefont
  {Campisi}}, \bibinfo {author} {\bibfnamefont {P.}~\bibnamefont {H\"{a}nggi}},
  \ and\ \bibinfo {author} {\bibfnamefont {P.}~\bibnamefont {Talkner}},\
  }\href@noop {} {\bibfield  {journal} {\bibinfo  {journal} {Rev. Mod. Phys.}\
  }\textbf {\bibinfo {volume} {83}},\ \bibinfo {pages} {771} (\bibinfo {year}
  {2011})}\BibitemShut {NoStop}%
\bibitem [{\citenamefont {M.~Esposito}\ and\ \citenamefont
  {Mukamel}(2009)}]{esposito}%
  \BibitemOpen
  \bibfield  {author} {\bibinfo {author} {\bibfnamefont {U.~H.}\ \bibnamefont
  {M.~Esposito}}\ and\ \bibinfo {author} {\bibfnamefont {S.}~\bibnamefont
  {Mukamel}},\ }\href@noop {} {\bibfield  {journal} {\bibinfo  {journal} {Rev.
  Mod. Phys.}\ }\textbf {\bibinfo {volume} {81}},\ \bibinfo {pages} {1665}
  (\bibinfo {year} {2009})}\BibitemShut {NoStop}%
\bibitem [{\citenamefont {del Rio}\ \emph {et~al.}(2011)\citenamefont {del
  Rio}, \citenamefont {Aberg}, \citenamefont {Renner}, \citenamefont
  {Dahlsten},\ and\ \citenamefont {Vedral}}]{delRio2011}%
  \BibitemOpen
  \bibfield  {author} {\bibinfo {author} {\bibfnamefont {L.}~\bibnamefont {del
  Rio}}, \bibinfo {author} {\bibfnamefont {J.}~\bibnamefont {Aberg}}, \bibinfo
  {author} {\bibfnamefont {R.}~\bibnamefont {Renner}}, \bibinfo {author}
  {\bibfnamefont {O.}~\bibnamefont {Dahlsten}}, \ and\ \bibinfo {author}
  {\bibfnamefont {V.}~\bibnamefont {Vedral}},\ }\href@noop {} {\bibfield
  {journal} {\bibinfo  {journal} {Nature}\ }\textbf {\bibinfo {volume} {474}},\
  \bibinfo {pages} {61} (\bibinfo {year} {2011})}\BibitemShut {NoStop}%
\bibitem [{\citenamefont {Dahlsten}\ \emph {et~al.}(2011)\citenamefont
  {Dahlsten}, \citenamefont {Renner}, \citenamefont {Rieper},\ and\
  \citenamefont {Vedral}}]{Dahlsten2011}%
  \BibitemOpen
  \bibfield  {author} {\bibinfo {author} {\bibfnamefont {O.}~\bibnamefont
  {Dahlsten}}, \bibinfo {author} {\bibfnamefont {R.}~\bibnamefont {Renner}},
  \bibinfo {author} {\bibfnamefont {E.}~\bibnamefont {Rieper}}, \ and\ \bibinfo
  {author} {\bibfnamefont {V.}~\bibnamefont {Vedral}},\ }\href@noop {}
  {\bibfield  {journal} {\bibinfo  {journal} {New J. Phys.}\ }\textbf {\bibinfo
  {volume} {13}},\ \bibinfo {pages} {053015} (\bibinfo {year}
  {2011})}\BibitemShut {NoStop}%
\bibitem [{\citenamefont {Brandao}\ \emph {et~al.}(2013)\citenamefont
  {Brandao}, \citenamefont {Horodecki}, \citenamefont {Oppenheim},
  \citenamefont {Renes},\ and\ \citenamefont {Spekkens}}]{Brandao2011}%
  \BibitemOpen
  \bibfield  {author} {\bibinfo {author} {\bibfnamefont {F.~G.~S.~L.}\
  \bibnamefont {Brandao}}, \bibinfo {author} {\bibfnamefont {M.}~\bibnamefont
  {Horodecki}}, \bibinfo {author} {\bibfnamefont {J.}~\bibnamefont
  {Oppenheim}}, \bibinfo {author} {\bibfnamefont {J.}~\bibnamefont {Renes}}, \
  and\ \bibinfo {author} {\bibfnamefont {R.~W.}\ \bibnamefont {Spekkens}},\
  }\href@noop {} {\bibfield  {journal} {\bibinfo  {journal} {Phys. Rev. Lett.}\
  }\textbf {\bibinfo {volume} {111}},\ \bibinfo {pages} {250404} (\bibinfo
  {year} {2013})}\BibitemShut {NoStop}%
\bibitem [{\citenamefont {Horodecki}\ and\ \citenamefont
  {Oppenheim}(2013)}]{Horodecki2013}%
  \BibitemOpen
  \bibfield  {author} {\bibinfo {author} {\bibfnamefont {M.}~\bibnamefont
  {Horodecki}}\ and\ \bibinfo {author} {\bibfnamefont {J.}~\bibnamefont
  {Oppenheim}},\ }\href@noop {} {\bibfield  {journal} {\bibinfo  {journal}
  {Nat. Commun.}\ }\textbf {\bibinfo {volume} {4}},\ \bibinfo {pages} {2059}
  (\bibinfo {year} {2013})}\BibitemShut {NoStop}%
\bibitem [{\citenamefont {Aberg}(2013)}]{Aberg2013}%
  \BibitemOpen
  \bibfield  {author} {\bibinfo {author} {\bibfnamefont {J.}~\bibnamefont
  {Aberg}},\ }\href@noop {} {\bibfield  {journal} {\bibinfo  {journal} {Nat.
  Commun.}\ }\textbf {\bibinfo {volume} {4}},\ \bibinfo {pages} {1925}
  (\bibinfo {year} {2013})}\BibitemShut {NoStop}%
\bibitem [{\citenamefont {Gour}\ \emph {et~al.}(2013)\citenamefont {Gour},
  \citenamefont {M\"uller}, \citenamefont {Narasimhachar}, \citenamefont
  {Spekkens},\ and\ \citenamefont {Halpern}}]{gour}%
  \BibitemOpen
  \bibfield  {author} {\bibinfo {author} {\bibfnamefont {G.}~\bibnamefont
  {Gour}}, \bibinfo {author} {\bibfnamefont {M.~P.}\ \bibnamefont {M\"uller}},
  \bibinfo {author} {\bibfnamefont {V.}~\bibnamefont {Narasimhachar}}, \bibinfo
  {author} {\bibfnamefont {R.~W.}\ \bibnamefont {Spekkens}}, \ and\ \bibinfo
  {author} {\bibfnamefont {N.}~\bibnamefont {Halpern}},\ }\href@noop {}
  {\bibfield  {journal} {\bibinfo  {journal} {arXiv:1309.6586}\ } (\bibinfo
  {year} {2013})}\BibitemShut {NoStop}%
\bibitem [{\citenamefont {Goold}\ \emph {et~al.}(2015)\citenamefont {Goold},
  \citenamefont {Paternostro},\ and\ \citenamefont {Modi}}]{landauer}%
  \BibitemOpen
  \bibfield  {author} {\bibinfo {author} {\bibfnamefont {J.}~\bibnamefont
  {Goold}}, \bibinfo {author} {\bibfnamefont {M.}~\bibnamefont {Paternostro}},
  \ and\ \bibinfo {author} {\bibfnamefont {K.}~\bibnamefont {Modi}},\
  }\href@noop {} {\bibfield  {journal} {\bibinfo  {journal} {Phys.~Rev.~Let.}\
  }\textbf {\bibinfo {volume} {114}},\ \bibinfo {pages} {060602} (\bibinfo
  {year} {2015})}\BibitemShut {NoStop}%
\bibitem [{\citenamefont {Allahverdyan}\ \emph {et~al.}(2004)\citenamefont
  {Allahverdyan}, \citenamefont {Balian},\ and\ \citenamefont
  {Nieuwenhuizen}}]{Allahverdyan}%
  \BibitemOpen
  \bibfield  {author} {\bibinfo {author} {\bibfnamefont {A.}~\bibnamefont
  {Allahverdyan}}, \bibinfo {author} {\bibfnamefont {R.}~\bibnamefont
  {Balian}}, \ and\ \bibinfo {author} {\bibfnamefont {T.}~\bibnamefont
  {Nieuwenhuizen}},\ }\href@noop {} {\bibfield  {journal} {\bibinfo  {journal}
  {Europhysis Letters}\ }\textbf {\bibinfo {volume} {67}},\ \bibinfo {pages}
  {565} (\bibinfo {year} {2004})}\BibitemShut {NoStop}%
\bibitem [{\citenamefont {Alicki}(1979)}]{Alicki1979}%
  \BibitemOpen
  \bibfield  {author} {\bibinfo {author} {\bibfnamefont {R.}~\bibnamefont
  {Alicki}},\ }\href@noop {} {\bibfield  {journal} {\bibinfo  {journal} {J.
  Phys. A: Math. Theor.}\ }\textbf {\bibinfo {volume} {12}},\ \bibinfo {pages}
  {L103} (\bibinfo {year} {1979})}\BibitemShut {NoStop}%
\bibitem [{\citenamefont {Spohn}(1978)}]{Spohn}%
  \BibitemOpen
  \bibfield  {author} {\bibinfo {author} {\bibfnamefont {H.}~\bibnamefont
  {Spohn}},\ }\href@noop {} {\bibfield  {journal} {\bibinfo  {journal} {J.
  Math. Phys.}\ }\textbf {\bibinfo {volume} {19}},\ \bibinfo {pages} {1227}
  (\bibinfo {year} {1978})}\BibitemShut {NoStop}%
\bibitem [{\citenamefont {Hatano}\ and\ \citenamefont
  {Sasa}(2001)}]{hatano2001steady}%
  \BibitemOpen
  \bibfield  {author} {\bibinfo {author} {\bibfnamefont {T.}~\bibnamefont
  {Hatano}}\ and\ \bibinfo {author} {\bibfnamefont {S.-I.}\ \bibnamefont
  {Sasa}},\ }\href@noop {} {\bibfield  {journal} {\bibinfo  {journal} {Phys.
  Rev. Lett.}\ }\textbf {\bibinfo {volume} {86}},\ \bibinfo {pages} {3463}
  (\bibinfo {year} {2001})}\BibitemShut {NoStop}%
\bibitem [{\citenamefont {Sagawa}(2012)}]{sagawa2013second}%
  \BibitemOpen
  \bibfield  {author} {\bibinfo {author} {\bibfnamefont {T.}~\bibnamefont
  {Sagawa}},\ }\href@noop {} {\bibfield  {journal} {\bibinfo  {journal}
  {Lectures on Quantum Computing, Thermodynamics and Statistical Physics}\
  }\textbf {\bibinfo {volume} {8}},\ \bibinfo {pages} {125} (\bibinfo {year}
  {2012})}\BibitemShut {NoStop}%
\bibitem [{\citenamefont {Janzing}\ \emph {et~al.}(2000)\citenamefont
  {Janzing}, \citenamefont {Wocjan}, \citenamefont {Zeier}, \citenamefont
  {Geiss},\ and\ \citenamefont {Beth}}]{Janzing}%
  \BibitemOpen
  \bibfield  {author} {\bibinfo {author} {\bibfnamefont {D.}~\bibnamefont
  {Janzing}}, \bibinfo {author} {\bibfnamefont {P.}~\bibnamefont {Wocjan}},
  \bibinfo {author} {\bibfnamefont {R.}~\bibnamefont {Zeier}}, \bibinfo
  {author} {\bibfnamefont {R.}~\bibnamefont {Geiss}}, \ and\ \bibinfo {author}
  {\bibfnamefont {T.}~\bibnamefont {Beth}},\ }\href@noop {} {\bibfield
  {journal} {\bibinfo  {journal} {International Journal of Theoretical
  Physics}\ }\textbf {\bibinfo {volume} {39}},\ \bibinfo {pages} {2717}
  (\bibinfo {year} {2000})}\BibitemShut {NoStop}%
\bibitem [{\citenamefont {Lenard}(1978)}]{lenard}%
  \BibitemOpen
  \bibfield  {author} {\bibinfo {author} {\bibfnamefont {A.}~\bibnamefont
  {Lenard}},\ }\href@noop {} {\bibfield  {journal} {\bibinfo  {journal} {J.
  Stat. Phys.}\ }\textbf {\bibinfo {volume} {19}},\ \bibinfo {pages} {575}
  (\bibinfo {year} {1978})}\BibitemShut {NoStop}%
\bibitem [{\citenamefont {Pusz}\ and\ \citenamefont {Woronowicz}(1978)}]{pusz}%
  \BibitemOpen
  \bibfield  {author} {\bibinfo {author} {\bibfnamefont {W.}~\bibnamefont
  {Pusz}}\ and\ \bibinfo {author} {\bibfnamefont {S.~L.}\ \bibnamefont
  {Woronowicz}},\ }\href@noop {} {\bibfield  {journal} {\bibinfo  {journal}
  {Comm. Math. Phys.}\ }\textbf {\bibinfo {volume} {58}},\ \bibinfo {pages}
  {273} (\bibinfo {year} {1978})}\BibitemShut {NoStop}%
\bibitem [{Note1()}]{Note1}%
  \BibitemOpen
  \bibinfo {note} {Since a passive state's internal energy is fixed even in the
  presence of degeneracies they have no bearing on the arguments presented here
  and the reader may think of passive states as being unique with regard to a
  given Hamiltonian: $\pi $ will here correspond to $H$, $\pi '$ to
  $H'$.}\BibitemShut {Stop}%
\bibitem [{\citenamefont {Hovhannisyan}\ \emph {et~al.}(2013)\citenamefont
  {Hovhannisyan}, \citenamefont {Perarnau-Llobet}, \citenamefont {Huber},\ and\
  \citenamefont {Ac\'in}}]{Hovhannisyan}%
  \BibitemOpen
  \bibfield  {author} {\bibinfo {author} {\bibfnamefont {K.~V.}\ \bibnamefont
  {Hovhannisyan}}, \bibinfo {author} {\bibfnamefont {M.}~\bibnamefont
  {Perarnau-Llobet}}, \bibinfo {author} {\bibfnamefont {M.}~\bibnamefont
  {Huber}}, \ and\ \bibinfo {author} {\bibfnamefont {A.}~\bibnamefont
  {Ac\'in}},\ }\href@noop {} {\bibfield  {journal} {\bibinfo  {journal} {Phys.
  Rev. Lett.}\ }\textbf {\bibinfo {volume} {111}},\ \bibinfo {pages} {240401}
  (\bibinfo {year} {2013})}\BibitemShut {NoStop}%
\bibitem [{\citenamefont {Alicki}\ and\ \citenamefont {Fannes}(2013)}]{fannes}%
  \BibitemOpen
  \bibfield  {author} {\bibinfo {author} {\bibfnamefont {R.}~\bibnamefont
  {Alicki}}\ and\ \bibinfo {author} {\bibfnamefont {M.}~\bibnamefont
  {Fannes}},\ }\href@noop {} {\bibfield  {journal} {\bibinfo  {journal} {Phys.
  Rev. E.}\ }\textbf {\bibinfo {volume} {87}},\ \bibinfo {pages} {042123}
  (\bibinfo {year} {2013})}\BibitemShut {NoStop}%
\bibitem [{\citenamefont {Giorgi}\ and\ \citenamefont
  {Campbell}(2015)}]{Campbell}%
  \BibitemOpen
  \bibfield  {author} {\bibinfo {author} {\bibfnamefont {G.~L.}\ \bibnamefont
  {Giorgi}}\ and\ \bibinfo {author} {\bibfnamefont {S.}~\bibnamefont
  {Campbell}},\ }\href@noop {} {\bibfield  {journal} {\bibinfo  {journal} {J.
  Phys. B: At. Mol. Opt. Phys.}\ }\textbf {\bibinfo {volume} {48}},\ \bibinfo
  {pages} {03550} (\bibinfo {year} {2015})}\BibitemShut {NoStop}%
\bibitem [{\citenamefont {Plastina}\ \emph {et~al.}(2014)\citenamefont
  {Plastina}, \citenamefont {Alecce}, \citenamefont {Apollaro}, \citenamefont
  {Falcone}, \citenamefont {Francica}, \citenamefont {Galve}, \citenamefont
  {{Lo Gullo}},\ and\ \citenamefont {Zambrini}}]{Plastina}%
  \BibitemOpen
  \bibfield  {author} {\bibinfo {author} {\bibfnamefont {F.}~\bibnamefont
  {Plastina}}, \bibinfo {author} {\bibfnamefont {A.}~\bibnamefont {Alecce}},
  \bibinfo {author} {\bibfnamefont {T.~J.~G.}\ \bibnamefont {Apollaro}},
  \bibinfo {author} {\bibfnamefont {G.}~\bibnamefont {Falcone}}, \bibinfo
  {author} {\bibfnamefont {G.}~\bibnamefont {Francica}}, \bibinfo {author}
  {\bibfnamefont {F.}~\bibnamefont {Galve}}, \bibinfo {author} {\bibfnamefont
  {N.}~\bibnamefont {{Lo Gullo}}}, \ and\ \bibinfo {author} {\bibfnamefont
  {R.}~\bibnamefont {Zambrini}},\ }\href@noop {} {\bibfield  {journal}
  {\bibinfo  {journal} {Phys. Rev. Lett.}\ }\textbf {\bibinfo {volume} {113}},\
  \bibinfo {pages} {260601} (\bibinfo {year} {2014})}\BibitemShut {NoStop}%
\bibitem [{\citenamefont {Vedral}(2002)}]{vedral2002role}%
  \BibitemOpen
  \bibfield  {author} {\bibinfo {author} {\bibfnamefont {V.}~\bibnamefont
  {Vedral}},\ }\href@noop {} {\bibfield  {journal} {\bibinfo  {journal} {Rev.
  Mod. Phys.}\ }\textbf {\bibinfo {volume} {74}},\ \bibinfo {pages} {197}
  (\bibinfo {year} {2002})}\BibitemShut {NoStop}%
\bibitem [{\citenamefont {Lindblad}(1975)}]{lindblad1975completely}%
  \BibitemOpen
  \bibfield  {author} {\bibinfo {author} {\bibfnamefont {G.}~\bibnamefont
  {Lindblad}},\ }\href@noop {} {\bibfield  {journal} {\bibinfo  {journal}
  {Communications in Mathematical Physics}\ }\textbf {\bibinfo {volume} {40}},\
  \bibinfo {pages} {147} (\bibinfo {year} {1975})}\BibitemShut {NoStop}%
\bibitem [{\citenamefont {Vinjanampathy}\ and\ \citenamefont
  {Modi}(2014)}]{secondLaw}%
  \BibitemOpen
  \bibfield  {author} {\bibinfo {author} {\bibfnamefont {S.}~\bibnamefont
  {Vinjanampathy}}\ and\ \bibinfo {author} {\bibfnamefont {K.}~\bibnamefont
  {Modi}},\ }\href@noop {} {\bibfield  {journal} {\bibinfo  {journal}
  {arXiv:1405.6140}\ } (\bibinfo {year} {2014})}\BibitemShut {NoStop}%
\bibitem [{\citenamefont {Faist}\ \emph {et~al.}(2014)\citenamefont {Faist},
  \citenamefont {Oppenheim},\ and\ \citenamefont {Renner}}]{faist}%
  \BibitemOpen
  \bibfield  {author} {\bibinfo {author} {\bibfnamefont {P.}~\bibnamefont
  {Faist}}, \bibinfo {author} {\bibfnamefont {J.}~\bibnamefont {Oppenheim}}, \
  and\ \bibinfo {author} {\bibfnamefont {R.}~\bibnamefont {Renner}},\
  }\href@noop {} {\bibfield  {journal} {\bibinfo  {journal} {arXiv:1406.3618}\
  } (\bibinfo {year} {2014})}\BibitemShut {NoStop}%
\bibitem [{Note2()}]{Note2}%
  \BibitemOpen
  \bibinfo {note} {In the same way, adiabatic work is positive when $H' \succ
  H$. Unlike for operational heat, there is, however, no connection between the
  majorisation relations for $H$ and $H'$ and the type of map that governs the
  process.}\BibitemShut {Stop}%
\bibitem [{\citenamefont {Pucha{\l}a}\ \emph {et~al.}(2013)\citenamefont
  {Pucha{\l}a}, \citenamefont {Rudnicki},\ and\ \citenamefont
  {{\.Z}yczkowski}}]{Zyczkowski}%
  \BibitemOpen
  \bibfield  {author} {\bibinfo {author} {\bibfnamefont {Z.}~\bibnamefont
  {Pucha{\l}a}}, \bibinfo {author} {\bibfnamefont {{\L}.}~\bibnamefont
  {Rudnicki}}, \ and\ \bibinfo {author} {\bibfnamefont {K.}~\bibnamefont
  {{\.Z}yczkowski}},\ }\href@noop {} {\bibfield  {journal} {\bibinfo  {journal}
  {J. Phys. A: Math. Theor.}\ }\textbf {\bibinfo {volume} {46}},\ \bibinfo
  {pages} {272002} (\bibinfo {year} {2013})}\BibitemShut {NoStop}%
\bibitem [{\citenamefont {Bengtson}\ and\ \citenamefont
  {Zyczkowski}(2006)}]{Bengtson}%
  \BibitemOpen
  \bibfield  {author} {\bibinfo {author} {\bibfnamefont {I.}~\bibnamefont
  {Bengtson}}\ and\ \bibinfo {author} {\bibfnamefont {C.}~\bibnamefont
  {Zyczkowski}},\ }\href@noop {} {\emph {\bibinfo {title} {Geometry of Quantum
  States}}}\ (\bibinfo  {publisher} {Cambridge University Press, Cambridge},\
  \bibinfo {year} {2006})\BibitemShut {NoStop}%
\end{thebibliography}%
\end{document}